\documentclass{elsart}
\usepackage[square]{natbib}
\usepackage[dvips]{graphics}
\begin{document}
\runauthor{Gong, Zejda, D\"appen and Aparicio}
\date{2000/09/06}
\begin{frontmatter}
\title{Generalized Fermi-Dirac Functions and Derivatives: 
Properties and Evaluation}
\author[Gong]{Zhigang Gong}
\author[Gong]{Ladislav Zejda}
\author[Gong]{Werner D\"appen}
\author[Aparicio]{Josep M. Aparicio}

\address[Gong]{Department of Physics and Astronomy, University of Southern
California, Los Angeles, CA 90089, U.S.A}
\address[Aparicio]{Earth Sciences Department , Institut d'Estudis Espacials 
de Catalunya, E-08034 Barcelona, Spain}
\begin{abstract}

The generalized Fermi-Dirac functions and their derivatives are important
in evaluating the thermodynamic quantities of partially degenerate electrons 
in hot dense stellar plasmas. New recursion relations of the 
generalized Fermi-Dirac functions have been found. An 
effective numerical method to evaluate the derivatives of the generalized 
Fermi-Dirac functions up to third order with respect to both 
degeneracy and temperature is then proposed, following Aparicio 
\citep{Apa}.
A Fortran program based on this method, together 
with a sample test case, is provided. Accuracy and domain of reliability
of some other, popularly used analytic approximations of the generalized
Fermi-Dirac functions for extreme conditions are investigated and
compared with our results.

\end{abstract}
\end{frontmatter}

{\bf PROGRAM SUMMARY}

{\it Title of program:} GFD\_D3

{\it Catalogue identifier:} ???? 

{\it Program Summary URL:} http://www.cpc.cs.qub.ac.uk/summaries/????

{\it Program obtainable from:} CPC Program Library, Queen's University
of Belfast, N. Ireland

{\it Licensing provisions:} none

{\it Computers:} Sun E4500/E5500, Compaq DEC Alpha, SGI Origin2000,
HP Convex Exemplar, Cray SV1-1A/16-8, AMD K6 PC, IBM SP2

{\it Operating systems under which the program has been tested:}
Solaris 5.6, Linux (Red Hat 5.2), IRIX 64, SPP-UX 5.3, Unicos 
10.0.0.6, Microsoft Windows 98 (2nd ed.), AIX 4.2 

{\it Programming language:} FORTRAN 77

{\it Number of bytes in distributed program, including test data, etc.:}
20758

{\it Distribution format:} uuencoded (gzip) compressed tar file


{\it Nature of physical problem\\}
Provide numerical method to evaluate generalized Fermi-Dirac functions
and their derivatives with respect to $\eta$ and $\beta$ up to third
order. The results are important for a highly accurate calculation of 
thermodynamic quantities 
of an electron gas with partial degeneracy and relatively high temperatures
with very high order of accuracy.

{\it Method of solution\\}
Following the scheme proposed by Aparicio \cite{Apa}, the generalized Fermi 
- Dirac integration is split into four optimized regions. Gauss-Legendre
quadrature is used in the first three pieces, and Gauss-Laguerre 
quadrature in the last part when the $\e^{-x}$ term in the 
integrand dominates. Different break points are individually
chosen for each $\eta$ 
derivative.

{\it Typical running time\\}
Less than 1 ms for each data point on a DEC Alpha station with a 533 MHz
CPU in double precision.

{\bf \large LONG WRITE-UP}

\section{Introduction}
\typeout{SET RUN AUTHOR to \@runauthor}

Relativistic quantum statistics is important for studying thermodynamic 
properties of electrons at the elevated degeneracy and relatively high 
temperature as found in the interior of massive 
stars. Chandrasekhar \cite{Chandra} expresses the 
thermodynamic quantities as integrals of hyperbolic functions [see 
Eq. (170) - (172) of \cite{Chandra}]. In a more 
general way Cox \& Giuli \cite{Cox}, 
write them in term of the generalized Fermi-Dirac (FD) 
functions with different orders [see Eq. (24.98) - (24.100) of
\cite{Cox}]. They define 
the generalized FD functions as

\begin{equation}\label{fo:gFD}
   F_k (\eta, \beta) = \int_0^\infty {{{x^k \sqrt {1+({\beta x / 2})}}
   \over {\e^{x-\eta} + 1}}} \d x \ ,
\end{equation}

where $\eta \equiv {\mu / k_{\mathrm B} T}$ and $\beta \equiv 
{k_{\mathrm B} T / m_e c^2}$ are the degeneracy parameter and a
dimensionless temperature,
respectively. Here, $m_e$ is electron mass, $T$ absolute temperature, 
$\mu$ the chemical potential, $c$ speed of light in vacuum and $k_{\mathrm 
B}$ the Boltzmann constant. The non-relativistic form of Eq.~(1)
(corresponding to $\beta = 0$) is

\begin{equation}\label{fo:FD}
   F_k (\eta) = \int_0^\infty {x^k \over {\e^{x-\eta} + 1}} \d x
\end{equation}

is usually called the Fermi-Dirac function, and is used to study 
thermodynamic properties of the degenerate non-relativistic electron gas. 
For the usual physical applications, both for the non-relativistic and
relativistic functions, the 
index $k$ is an integer or half integer with $k \geq -1$.

No exact analytic formulae are available to evaluate
the generalized FD functions in full range of degeneracy and temperature.
Some approximate expressions exist for some extreme cases, such
as when degeneracy is very low, or extremely high \cite{Cox, Mira, Pichon,
Blud}. There are also asymptotic expressions for the thermodynamic
quantities alone \cite{Blin}. On the other hand, Eggleton et al. \cite{Egg},  
give a very useful global fitting formula for the generalized FD functions 
[see also \cite{Pichon}], even though its accuracy is relatively modest.

Numerical evaluation of the generalized FD functions is
not an easy task. Simple tabular interpolations \cite{Cox} can cover 
only a small range of degeneracy and temperature, and the accuracy is not
very good. Because the integrand is proportional 
to $x^k$ when $x \rightarrow 0$, and decays 
as $\e^{-x}$ when $x \rightarrow \infty$, few simple numeric 
integration methods can simultaneously cover a sufficiently large range of 
degeneracy and temperature with acceptable accuracy as well as with an
acceptable calculation effort. \cite{Sagar} uses a modified Gauss quadrature
with weight function $w(x)=x^k/(\e^x+1)$, and \cite{Gau}
uses weight function $w(x)=x^k/\e^x$ to integrate the generalized 
FD functions directly. Both methods are relatively fast, but 
are only good for $\eta \leq 5$. In \cite{Pichon}, a Gauss-Laguerre 
quadrature with adjusted decay factor is used to cover $-1 \leq \eta 
\leq 11$, and approximation formulae beyond that region. In
\cite{Mira, NaMo}, different variations of the trapezoidal rule are
being used to do the integration, 
but both involve a heavy trade-off between accuracy and
computing time.

Mohankumar \& Natarajan \cite{MoNa} use another approximation method by 
subtracting the sum of residues of finite number of poles within a pre-selected 
contour from results of trapezoidal integration. This method is fast when 
$\eta$ is not too big, and the results can reach 14-digit accuracy in 
double precision for $\beta < 100$, and when $\eta$ is not too small, 
typically 
{}from $\eta > -25$ to $\eta =2000$. It is good to about 10 digits when $\eta$ 
is as large as 10000. However, the number of poles to be included increases 
rapidly when $\eta$ gets too big. Extending this method to calculate different
derivatives of the generalized FD functions with the same order of accuracy is 
not an easy task.

Aparicio \cite{Apa} proposed another method 
which consists of splitting the generalized-FD-function 
integral into four regions. The break points are chosen such 
that errors from all four regions are balanced, and the location of the 
break points is optimized. Based on the property of the integrand, the
Gauss-Legendre quadratures used are from $x=0$ to $x=s_3$, and the 
Gauss-Laguerre quadrature from $x=s_3$ to $x \rightarrow \infty$
[see Appendix (\ref{BPs})]. The method is fast (proportional to the number 
of points used in Gauss quadratures) and accurate. Extending it to 
include derivatives of the generalized FD functions is straightforward. 
Our numerical method to evaluate derivatives of the 
generalized FD functions follows this scheme closely. 
We discuss it in more detail in Sect.~3.

Besides the generalized FD functions themselves, their derivatives are also 
important for computing the thermodynamic quantities of an electron gas. 
Miralles \& Van Riper \cite{Mira} give the results of the first 
order of derivatives
of the generalized FD functions with respect to both $\eta$ and $\beta$.
They use approximation formulae when $\eta$ is either very large or very
small, and interpolate through tables when $\eta$ is in between. The
disadvantage is that it is only accurate to about 6 digits, and discontinuity
exists when two different methods join. Also, their results are limited to
$k= {1\over2}, {3\over2}$ and ${5\over2}$ only. On the other
hand, solar equation of state studies require up to third order derivatives
on the generalized FD functions \cite{CD92}. In particular, 
the precise data from helioseismology
have demonstrated that the relativistic electron
gas is necessary for solar modeling \cite{EK98}. For solar applications,
high-precision computations of the relativistic thermodynamic quantities
are mandatory \cite{Gong}.

Because of its complexity, the generalized FD functions have not been fully
understood. In Sect. 2 we first
present some newly found recursion relations, and then
discuss some of their properties. The numerical method to calculate 
derivatives of the generalized FD functions up to the third order is 
introduced in Sect. 3, based on the scheme of \cite{Apa}. At the end we 
check the efficiency and validity of some popularly used analytic
approximation formulae in several extreme cases in Sect. 4.

\section{Recursion Relations}

Cox \& Giuli \cite{Cox} show that the generalized FD functions obey the 
following recursion relations

\begin{equation}\label{fo:RR1}
   \beta {\partial F_k (\eta, \beta) \over \partial \beta} = 
   {\partial F_{k+1}  (\eta, \beta) \over \partial \eta} -
   (k+1) F_k (\eta, \beta)
\end{equation}
which link the FD functions $F_k (\eta, \beta)$ and $ F_{k+1}  (\eta, 
\beta)$ {\it via} both their $\eta$ and $\beta$ derivatives. Beyond 
that, we have found that the generalized FD functions also obey another 
recursion relation

\begin{equation}\label{fo:RR2}
   F_{k+1}  (\eta, \beta) = 4 \; {\partial F_k (\eta, \beta) \over 
   \partial \beta} + 2 \beta \; {\partial F_{k+1}  (\eta, \beta) \over
   \partial \beta} \ ,
\end{equation}
which is between the generalized FD functions and their $\beta$ derivatives.

{}From Eq. (\ref{fo:RR1}) and Eq. (\ref{fo:RR2}) we can further derive 
two useful recursion relations, in which

\begin{equation}\label{fo:RR3}
   {\partial F_{k+1}  (\eta, \beta) \over \partial \eta} = \left(k+{3 \over 2}
   \right) F_k (\eta, \beta) - 2\; {\partial F_{k-1}  (\eta, \beta) \over 
   \partial \beta}
\end{equation}
is a relation among $F_{k-1}  (\eta, \beta)$, $F_k (\eta, \beta)$ and $ F_{k+1}
(\eta, \beta)$, with both $\eta$ and $\beta$ derivatives, and

\begin{equation}\label{fo:RR4}
   k F_{k-1}  (\eta, \beta) + \left( {1\over2}\:k + {3\over4} \right) \beta 
   F_k (\eta, \beta) = {\partial F_k (\eta, \beta) \over \partial \eta} + 
   {1\over2} \beta \; {\partial F_{k+1}  (\eta, \beta) \over \partial \eta}
\end{equation}
is between the generalized FD functions and their $\eta$ derivatives only.

These recursion relations can be used to check the accuracy of numerical
evaluations as well as to simplify some theoretical expressions.
However, one should keep in mind that when $\eta$ is very small the two
terms on the right-hand side of
either Eq. (\ref{fo:RR1}) and Eq. (\ref{fo:RR3}) are so close
to each other
that almost perfect cancellation happens. For this reason,
the use of
recursion relations to evaluate the derivatives of the generalized FD 
functions
is not always possible.

One interesting property of the recursion relation in Eq. (\ref{fo:RR1}) is
that, for the non-relativistic limit case ($\beta=0$), it
can be written as:

\begin{equation}\label{fo:RR5}
   {\partial F_{k+1}  (\eta) \over \partial \eta} = (k+1) F_k (\eta)
\end{equation}

which can be further written as:

\begin{equation}\label{fo:RR6}
   \int_0^\infty \frac{x^k}{\e^{x-\eta} + 1} \d x = \int_0^\infty 
\frac{x^k}{\e^{x-\eta} + 1} \; \frac{x}{(\e^{x-\eta} + 1)(k+1)} \d x
\end{equation}

Both Eqs. (\ref{fo:RR5}) and (\ref{fo:RR6}) are satisfied for arbitrary 
$\eta$. This is very remarkable, because although the function 
${x}/{[(\e^{x-\eta} + 1)(k+1)]}$ is not equal to $1$, 
when multiplied with the Fermi-Dirac kernel 
${x^k}/{(\e^{x-\eta} + 1)}$, it leads to exactly the same value of the
definitive integrals as if it were equal to $1$.

Another interesting behavior we observed from the generalized FD functions
is the dependence on $\eta$ of the following ratio (see Fig.
\ref{fig:Cancel}). 

\begin{equation}\label{fo:Cancel}
   \{ [\partial F_k (\eta, \beta) / \partial \eta] + \beta
   [\partial F_{k+1}  (\eta, \beta) / \partial \eta] \}^2 \over
   [F_k (\eta, \beta) + \beta F_{k+1}  (\eta, \beta)] \: \:
   \{ [\partial^2 F_k (\eta, \beta) / \partial \eta^2] + \beta
   [\partial^2 F_{k+1}  (\eta, \beta) / \partial \eta^2] \}
\end{equation}
The expression appears in the derivatives of number 
density or internal energy of the relativistic 
electron gas (see \cite{GD} for more detail).

\begin{figure}
\vspace{10mm}
\caption{ PLEASE PLACE FIGURE (\ref{fig:Cancel}) HERE.}
\label{fig:Cancel}
\end{figure}

For small degeneracy ($\eta \rightarrow -\infty$) the ratio 
(\ref{fo:Cancel}) is very close 
to 1 for all value of index $k$ ($k>-1$) and all $\beta \geq 0$. However, 
when the electron gas is extremely degenerate ($\eta 
\rightarrow \infty$), the ratio (\ref{fo:Cancel}) 
approaches different limits for the non-relativistic
and relativistic cases. For the non-relativistic case
($\beta=0$), the limits of (\ref{fo:Cancel}) 
are ${1\over -1}, {3\over1}, {5\over3}, {7\over5},
...$ when $k=-{1\over2},{1\over2}, {3\over2},{5\over2}, ...$, 
respectively. The general expression is ${k+1 \over k}$ 
for $k>-1$ and $k\neq 0$. For the relativistic case ($\beta \neq 0$) 
the limits of (\ref{fo:Cancel}) 
are ${2\over1}, {3\over2}, {4\over3}, {5\over4},...$ when 
$k=-{1\over2}, {1\over2}, {3\over2}, {5\over2},...$, respectively. The
general expression is ${k+5/2 \over k+3/2}$ ($k>-1$). 
This result is true for $\eta \: \beta \gg 1$. So for a given small 
$\beta$ ($\beta \neq 0$),
that is, in the nearly non-relativistic case, for sufficiently large
$\eta$, the relativistic relation is satisfied, as shown in Fig. 
(\ref{fig:Cancel}).
This property 
links the behavior of the generalized FD functions of both extreme cases
(very small degeneracy or extremely great degeneracy). In Appendix
(\ref{Ap:Cancel}) we will give a 
proof of this property using asymptotic expansions.

One thing we want to point out is that, because the ratio of Eq. 
(\ref{fo:Cancel}) is very close to unity when $\eta$ is small, one has to 
be cautious when evaluating the difference between the nominator and
denominator for the case of a slightly 
degenerate electron gas in order to avoid numerical noise due to near
cancellation.

\section{Numerical Method and Test Run Results}

We choose the scheme proposed by \cite{Apa} to calculate derivatives
of the generalized FD functions, because this method is fast, accurate and
transparent. The expressions for the derivatives are listed in Appendix 
(\ref{Derivative}). In the method adopted, 
the integration is further split into
four domains. Gauss-Legendre quadratures are applied to the first three
pieces, starting from $x=0$ to a preselected point $s_3$, as defined in 
Appendix (\ref{BPs}). In the last part, where $x$ is from $s_3$ to $\infty$
the $\e^{-x}$ term dominates, and hence Gauss-Laguerre quadrature is used. 
The positions of the break points are optimized such 
that the combined errors from all the pieces contributing to the 
generalized FD functions are minimized and equally distributed among
the pieces. Fortunately, 
the choice of break points turns out rather uncritical
for each individual derivative [see also \cite{Apa}].

As shown in Appendix (\ref{Derivative}), the integrand of each derivative 
can be represented as a product of the generalized FD kernel 
$\Big[x^k {\sqrt {1+(\beta x / 2)}}\Big]/{(\e^{x-\eta} + 1)}$ 
with a tuning function. 
When $\eta$ and $\beta$ are not very large, this tuning function varies mildly
[see Fig. (\ref{fig:TF})]. Hence the same break points chosen for 
the generalized 
FD functions themselves can also 
be applied to their derivatives, and good accuracy is
achieved. However, when $\eta$ and $\beta$ are getting large, new break
points must be chosen. 
The $\beta$ derivatives are not very sensitive to the
choice of break points. Therefore, we provide new break points only for $\eta$ 
derivatives. Formulae to calculate break points can be found in 
Appendix (\ref{BPs}).

\begin{figure}
\vspace{10mm}
\caption{ PLEASE PLACE FIGURE (\ref{fig:TF}) HERE.}
\label{fig:TF}
\end{figure}

One concern in the calculation of 
the generalized FD functions and their derivatives is
how one determines the accuracy of the results. 
For the generalized FD functions 
we compare the results of \cite{Apa} (the method adopted in the present 
paper) with those of 
\cite{MoNa} evaluated in double precision. 
We find that they agree with each other 
to 14-digit accuracy for $\eta$ up to 1000, and to 10-digit accuracy 
for $\eta$ up to
10000. Additionally, both of them agree with \cite{Mira} to 8-to-10-digit 
accuracy for $\eta$ up to 1000. 

For the derivatives of the generalized FD functions,
we first test them using the different recursion relations available. 
Except for the aforementioned cancellations when $\eta$ is small, 
the recursion 
relations are satisfied with better than 10-digit accuracy,
for all derivatives with
$\beta$ being between $10^{-6}$ and $10^4$ and $\eta$ between -100 and
10000.
The only exception is the
third-order derivative with respect to $\eta$. Here, the same 
accuracy can only be achieved for $\eta$ less than 500. All of the above
results refer to double precision (128 bits) (which is single precision
on a Cray). 
Of course, better results can be achieved with quadruple
precision. In practice, if $k \neq -{1\over 2}$, the recursion relations
can be used to convert the third-order $\eta$ derivative into a lower-order
one when $\eta$ is big, and in this way, very accurate results can be 
achieved.

As a second test, 
we compare our directly evaluated derivatives with those obtained from
numerical differentiation. For example, we compare ${\partial F_{1/2}
(\eta, \beta) / \partial \eta}$ from our method with $\Big[ {F_{1/2}
(\eta+\Delta \eta,  \beta) - F_{1/2}(\eta-\Delta \eta,  \beta)\Big] /
2\: \Delta \eta}$, with the difference $\Delta \eta$ 
chosen to be $10^{-2}$, $10^{-4}$, $10^{-6}$, $10^{-8}$, 
respectively, and the calculation is done 
in double precision. The purpose of the test
is (i) to test the accuracy 
of our method, and (ii) to find the dependence of $\Delta \eta$ as
a function of accuracy. 

We conclude that both methods agree well. In practice, a step of 
$\Delta \eta$ or $\Delta \beta$ between $10^{-2}$ and $10^{-4}$
is an optimal choice to compute the derivatives of the generalized FD
functions 
by numerical differentiation.
For example, the numerical evaluation of the
second-order derivative with respect to $\eta$
{}from the first-order derivative yields an accuracy of at least 8 digits
in double precision. 

However, we would like to stress that our direct evaluations are
superior to the numerical differentiation used here only for testing
purposes. The accuracy of numerical differentiation
can drop to half of the machine precision [{\it i.e.} to
8 to 9 digits, compared to the 16 digit accuracy of 
double precision (non-Cray machines) or single precision 
(Cray machines)] when the 
aforementioned 3-point numerical differentiation 
is applied once. Applied
twice, the resulting accuracy can drop to 3 to 5 digits for $\eta < 2000$, 
and only 2 digits when $\eta \approx 5000$. 
Applied three times,
the resulting accuracy would only be 2 
digits when $\eta \leq 50$, and for higher $\eta$ values, the 
results would be completely useless. This demonstrates the well-known
fact
that numerical differentiation is a dangerous procedure if it is used
beyond first-order derivatives.
In addition, the step size to be used has to be chosen judiciously.
For high-precision applications such as found in helioseismology,
which requires
precise second- and third-order thermodynamic quantities,
the direct approach method presented in this paper is by far superior 
to more-than-one-time numerical differentiations.

In this paper we have always used 200-point Gauss quadratures. Although
fewer-points quadratures are also good for the calculations of the 
generalized FD functions 
\cite{Apa}, the 200-point Gauss quadratures used here guarantee the
accuracy of the high-order derivatives.
If computing speed is important, and
third-order $\eta$-derivatives are not needed, then the 40-to-80-point 
Gauss quadratures in \cite{Apa} are a good choice.

To use the program accompanying 
this paper, the user has first to compile and link 
the three Fortran files.
Then, upon choosing 
the index of the requested
generalized FD function $k$, the label ${\rm IB}$ denoting either
the function itself or a specified derivative
(see Table \ref{tb:Index}), the range of degeneracy 
$\eta$ and dimensionless temperature $\beta$,
the program then calculates the desired generalized FD function
or
derivative. A sample input is given in 
the file {\it input}, and the corresponding 
results are contained in the file {\it testResult}.

\begin{table}
\vspace{10mm}
\caption{ PLEASE PLACE TABLE (\ref{tb:Index}) HERE.}
\label{tb:Index} 
\end{table}

\section{Validity of Analytic Approximations}

Although the present numerical method to evaluate
generalized FD functions is accurate within a very large range of input
parameters $\eta$ and
$\beta$, analytic approximations still have advantages for theoretical 
studies and the understanding of the underlying physics. However, since
analytic approximations are usually only valid for restricted parameter
regions, we must know them for each of them. In the following, we limit
ourselves to the properties of the $k={1\over2}$ functions only if its
behavior is representative for the other cases. Otherwise,
we discuss the cases of index $k={1\over2}$ and $k={3\over2}$.

First we check the accuracy of the non-relativistic (NR) FD function
as an approximation to the generalized FD functions in the weakly 
non-relativistic case ($\beta \ll 1$)

\begin{equation}\label{fo:NR}
   F_k (\eta, \beta) \simeq F_k (\eta)  \ .
\end{equation}

\begin{figure}
\vspace{10mm}
\caption{PLEASE PLACE FIGURE (\ref{fig:NR}) HERE.}
\label{fig:NR}
\end{figure}

{}From Fig. \ref{fig:NR}, we can see that when both $\eta$ and $\beta$ are 
small the approximation is good to 6 digits, and the accuracy drops when
either $\eta$ or $\beta$ is getting big. Considerably better agreement 
(twice as many accurate digits, as shown 
in Fig. \ref{fig:NR1}) can be reached by including 
one more term is \cite{Blud}:

\begin{equation}\label{fo:NR1}
   F_k (\eta, \beta) \simeq F_k (\eta) + {1\over4} \beta F_{k+1} (\eta)
\end{equation}

\begin{figure}
\vspace{10mm}
\caption{PLEASE PLACE FIGURE (\ref{fig:NR1}) HERE.}
\label{fig:NR1}
\end{figure}

For the ultra-relativistic case ($\beta \gg 1$) \cite{Blud}, results of the 
following approximation is shown in Fig. \ref{fig:ER}:

\begin{equation}\label{fo:ER}
   F_k (\eta, \beta) \simeq \sqrt{\beta \over 2}\: \left[F_{k+{1\over2}} (\eta)
   + {1\over \beta} F_{k-{1\over2}} (\eta) \right]
\end{equation}

The figure shows that more than 6-digit accuracy can be obtained
with this formula if
$\beta > 0$, and $\eta$ is not too small, typically $\eta \geq 0$, and
more than 10-digit accuracy when $\beta > 2$.

\begin{figure}
\vspace{10mm}
\caption{PLEASE PLACE FIGURE (\ref{fig:ER}) HERE.}
\label{fig:ER}
\end{figure}

When $\eta>0$ we can see that the asymptotic expansions by Pichon \cite{Pichon} 
[see Appendix C.1, Eq.~(C1-C11); also Cox \& Giuli \cite{Cox}] 
are very good for $F_{1\over2} (\eta, \beta)$
(Fig. \ref{fig:ED12}), with better than 12-digit accuracy when $\eta > 120$.
However, the results are only moderate for $F_{3\over2} (\eta, \beta) $ (Fig. 
\ref{fig:ED32}) and $F_{5\over2} (\eta, \beta) $, in which case the accuracy 
is only better than 6 digits when $\eta > 5000$, and much less accurate when
$\eta$ is smaller. For convenience, We list the formulae for large and
small $\eta$ in Appendix (\ref{Asymp}).

\begin{figure}
\vspace{10mm}
\caption{PLEASE PLACE FIGURE (\ref{fig:ED12}) HERE.}
\label{fig:ED12}
\end{figure}

\begin{figure}
\vspace{10mm}
\caption{PLEASE PLACE FIGURE (\ref{fig:ED32}) HERE.}
\label{fig:ED32}
\end{figure}

When $\eta$ is small, the expression of the generalized FD functions in terms
of modified Bessel functions [see Appendix C.2, Eq.~(C12-C17); 
Cox \& Giuli \cite{Cox}] achieve very high accuracy.
We have used Mathematica 4.0 (for Solaris) to evaluate the modified Bessel 
functions in double precision. Since the leading terms in Eq. (\ref{fo:SD1}) -
(\ref{fo:SD3}) are comparable when $\beta$ is small, cancellation effect
would appear. A power series expansion expression \cite{Mira} is
therefore more appropriate. We have combined the methods 
and show their results in Fig. \ref{fig:ND}.
 
\begin{figure}
\vspace{10mm}
\caption{PLEASE PLACE FIGURE (\ref{fig:ND}) HERE.}
\label{fig:ND}
\end{figure}

We have also checked the quality of the fitting formulae by \cite{Egg}
[see Appendix C.3, Eq.~(C21-C26)] (Fig. \ref{fig:Egg}). 
Specifically, we have adopted the expression given 
by \cite{Pichon} (his Eq.~5a - 5c). 
These fitting formulae cover a much larger range of $\eta$ and 
$\beta$ than those asymptotic formulae above (but with lower accuracy), 
and they are adequate if
accuracy is not critical, and temperature not
too high. For instance, 
with these fitting formulae
we can obtain 4-digit accuracy when $\eta < 1000$ and $\beta < 10^{-4}$,
and 2-digit accuracy when $\eta < 10000$ and $\beta < 3 \times 10^{-5}$.

\begin{figure}
\vspace{10mm}
\caption{PLEASE PLACE FIGURE (\ref{fig:Egg}) HERE.}
\label{fig:Egg}
\end{figure}

\section{Conclusions}

In this paper we have
investigated properties of the generalized FD functions in
some detail, and we have
found some useful new recursion relations. We
propose an accurate and relatively fast numerical method to evaluate the 
derivatives of the generalized FD functions, following the algorithm
introduced by \cite{Apa}. 
Limitations of direct numerical differentiation are also discussed.
Finally, we have obtained the
range of validity of some popular analytic approximations 
for the generalized FD functions. 

\ack
ZG would like to thank 
A. Natarajan and N. Mohabkumar for providing us with their code
to evaluate generalized FD functions, and A. Nayfonov for useful
discussions. ZG, LZ and WD are supported in part by the 
NSF grant AST-9618549 and AST-9987391.

\appendix
\section{APPENDIX: Derivatives of the Generalized FD Functions}
  \label{Derivative}

Derivatives of the generalized FD functions up to third order with respect to
$\eta$ and $\beta$.

\begin{equation}\label{fo:DR1}
\frac{\partial F_k(\eta, \beta)}{\partial \eta} = \int_0^\infty 
\frac{x^k \sqrt {1+{\beta x \over 2}}}{\e^{x-\eta} + 1} \: \frac{1}
{1+\e^{\eta-x}} \d x
\end{equation}

\begin{equation}\label{fo:DR2}
\frac{\partial F_k(\eta, \beta)}{\partial \beta} = \int_0^\infty 
\frac{x^k \sqrt {1+{\beta x \over 2}}}{\e^{x-\eta} + 1} \: \frac{x}
{4+2 \beta x} \d x
\end{equation}

\begin{equation}\label{fo:DR3}
\frac{\partial^2 F_k(\eta, \beta)}{\partial \eta^2} = \int_0^\infty 
\frac{x^k \sqrt {1+{\beta x \over 2}}}{\e^{x-\eta} + 1} \: \frac{1-\e^
{\eta-x}}{(1+\e^{\eta-x})^2} \d x
\end{equation}

\begin{equation}\label{fo:DR4}
\frac{\partial^2 F_k(\eta, \beta)}{\partial \eta \partial \beta} = \int_0^
\infty \frac{x^k \sqrt {1+{\beta x \over 2}}}{\e^{x-\eta} + 1} \: \frac{x}
{4+2 \beta x} \: \frac{1}{1+\e^{\eta-x}} \d x
\end{equation}

\begin{equation}\label{fo:DR5}
\frac{\partial^2 F_k(\eta, \beta)}{\partial \beta^2} = -\int_0^\infty 
\frac{x^k \sqrt {1+{\beta x \over 2}}}{\e^{x-\eta} + 1} \: \frac{x^2}
{(4+2 \beta x)^2} \d x
\end{equation}

\begin{equation}\label{fo:DR6}
\frac{\partial^3 F_k(\eta, \beta)}{\partial \eta^3} = \int_0^\infty 
\frac{x^k \sqrt {1+{\beta x \over 2}}}{\e^{x-\eta} + 1} \: \frac{(1-\e^
{\eta-x})^2-2\e^{\eta -x}}{(1+\e^{\eta-x})^3} \d x
\end{equation}

\begin{equation}\label{fo:DR7}
\frac{\partial^3 F_k(\eta, \beta)}{\partial \eta^2 \partial \beta} = \int_0^
\infty \frac{x^k \sqrt {1+{\beta x \over 2}}}{\e^{x-\eta} + 1} \: \frac{1-\e^
{\eta-x}}{(1+\e^{\eta-x})^2}  \: \frac{x} {4+2 \beta x}\d x
\end{equation}

\begin{equation}\label{fo:DR8}
\frac{\partial^3 F_k(\eta, \beta)}{\partial \eta \partial \beta^2} = -\int_0^
\infty \frac{x^k \sqrt {1+{\beta x \over 2}}}{\e^{x-\eta} + 1} \: \frac{x^2}
{(4+2 \beta x)^2} \: \frac{1} {1+\e^{\eta-x}}\d x
\end{equation}

\begin{equation}\label{fo:DR9}
\frac{\partial^3 F_k(\eta, \beta)}{\partial \beta^3} = \int_0^\infty 
\frac{x^k \sqrt {1+{\beta x \over 2}}}{\e^{x-\eta} + 1} \frac{3x^3}
{(4+2 \beta x)^3} \d x
\end{equation}

\section{APPENDIX: Break Points}
  \label{BPs}
Three break points $S_1, S_2, S_3$ are [as shown in \cite{Apa}]:
\begin{eqnarray}\label{fo:BP}
S_1 & = & X_a - X_b\\
S_2 & = & X_a\\
S_3 & = & X_a + X_c
\end{eqnarray}
where
\begin{equation}\label{fo:Xa}
X_a=\frac{a_1+b_1 \xi + c_1 \xi^2}{1+c_1 \xi}
\end{equation}

\begin{equation}\label{fo:Xb}
X_b=\frac{a_2+b_2 \xi + c_2 d_2 \xi^2}{1+e_2 \xi + c_2 \xi^2}
\end{equation}

\begin{equation}\label{fo:Xc}
X_c=\frac{a_3+b_3 \xi + c_3 d_3 \xi^2}{1+e_3 \xi + c_3 \xi^2}
\end{equation}

\begin{equation}\label{fo:Xi}
\xi (\eta) = \sigma^{-1} \ln [1+\e^{\sigma(\eta - D)}]
\end{equation}

where the corresponding parameters are:

\begin{center}
\begin{tabular}{|c llll|}
\hline \hline
Parameter & $F_k(\eta,\beta)$ & $\partial F_k(\eta,\beta) \over \partial \eta$
& $\partial^2 F_k(\eta,\beta) \over \partial \eta^2$ &  $\partial^3 F_k(\eta,
\beta) \over \partial \eta^3$\\
\hline 
$D$ &      3.3609E0 & 4.99551E0 & 3.93830E0 & 4.17444E0\\
$\sigma$ & 9.1186E-2 & 9.11856E-2 & 9.11856E-2 & 9.11856E-2\\
$a_1$ &    6.7774E0 & 6.77740E0 & 6.77740E0 & 6.77740E0\\
$b_1$ &    1.1418E0 & 1.14180E0 & 1.14180E0 & 1.14180E0\\
$c_1$ &    2.9826E0 & 2.98255E0 & 2.98255E0 & 2.98255E0\\
$a_2$ &    3.7601E0 & 3.76010E0 & 3.76010E0 & 3.76010E0\\
$b_2$ &    9.3719E-2 & 9.37188E-2 & 9.37188E-2 & 9.37188E-2\\
$c_2$ &    2.1064E-2 & 2.10635E-2 & 2.10635E-2 & 2.10635E-2\\
$d_2$ &    3.1084E+1 & 3.95015E1 & 3.14499E1 & 3.05412E1\\
$e_2$ &    1.0056E0 & 1.00557E0 & 1.00557E0 & 1.00557E0\\
$a_3$ &    7.5669E0 & 7.56690E0 & 7.56690E0 & 7.56690E0\\
$b_3$ &    1.1695E0 & 1.16953E0 & 1.16953E0 & 1.16953E0\\
$c_3$ &    7.5416E-1 & 7.54162E0 & 7.54162E0 & 7.54162E0\\
$d_3$ &    6.6559E0 & 7.64734E0 & 6.86346E0 & 7.88030E0\\
$e_3$ &   -1.2819E0 & -1.28190E-1 &-1.28190E-1 &-1.28190E-1\\
\hline 
\end{tabular}
\end{center}

\section{APPENDIX: Asymptotic Expressions for Great and Small Degeneracy}
  \label{Asymp}
\subsection{Great degeneracy ($\eta>0$) \cite{Cox, Pichon}}

\begin{eqnarray}\label{fo:ED1}
F_{1\over2} (\eta, \beta) & = & {1\over \sqrt{2\beta^3}}\: \left\{f_{1\over2}(y)
+(1+\eta \beta)[C_1+C_2+C_3(4y^2+7)]\right\}
\end{eqnarray}

\begin{eqnarray}\label{fo:ED2}
F_{3\over2} (\eta, \beta) & = & {1\over \sqrt{2\beta^5}}\: \Big\{f_{3\over2}(y)
+C_1(3+2\eta \beta)-C_2 \nonumber
\\& & -C_3[(4\eta \beta+6)\eta \beta+3] \Big\}
\end{eqnarray}

\begin{eqnarray}\label{fo:ED3}
F_{5\over2} (\eta, \beta) & = & {1\over \sqrt{2\beta^7}} \: \Big( 
f_{5\over2} (y) +C_1(5+\eta \beta) +C_2\{[(2\eta \beta+10)\eta \beta +15]
\eta \beta+5\} \nonumber
\\& & +C_3(3+5\eta \beta) \Big)
\end{eqnarray}

where
\begin{equation}
1+y^2=(1+\eta \beta)^2
\end{equation}

\begin{equation}
C_1={\pi^2 \over 6}\: {\beta^2 \over y}, \;\;\; C_2={7\over20}C_1\left({\pi
\beta \over y^2}\right)^2, \;\;\; C_3={31\over168}C_1\left({\pi \beta \over 
y^2}\right)^4
\end{equation}

For $\eta \geq 0.05$:

\begin{equation}
f_{1\over2}={1\over2}\left[y\sqrt{1+y^2}-\mathrm{arcsinh} (y)\right]
\end{equation}

\begin{equation}
f_{3\over2}={1\over3}y^3-f_{1\over2}(y)
\end{equation}

\begin{equation}
f_{5\over2}={5\over8}y\left(1+{2\over5}y^2\right)\sqrt{1+y^2}-{2\over3}
y^3-{5\over8}\mathrm{arcsinh} (y)
\end{equation}

For $\eta < 0.05$:

\begin{eqnarray}
f_{1\over2} & = & \Bigg( \bigg\{ \bigg[ \bigg( \Big\{ \Big[ \Big( -{58773
\over1114112} \: y^2+ {77 \over5120}\Big) \: y^2 -{63\over 3328} \Big] \: 
y^2+{35\over1408}\Big\} \: y^2-{5\over144} \bigg) \: y^2\nonumber
\\& &+{5\over36}\bigg] \: y^2-{1\over10}\bigg\} \: y^2+{1\over3} \Bigg) \: y^3
\end{eqnarray}

\begin{eqnarray}
f_{3\over2} & = & \bigg\{\bigg[\bigg(\Big\{\Big[\Big({58773\over1114112}
\: y^2-{77\over5120}\Big) \: y^2+{63\over 3328}\Big] \: y^2-{35\over1408}
\Big\} \: y^2+{5\over144}\bigg) \: y^2\nonumber
\\ & & -{5\over36}\bigg] \: y^2+{1\over10}\bigg\} \:  y^5
\end{eqnarray}

\begin{eqnarray}
f_{5\over2} & = & \bigg[\bigg(\Big\{\Big[\Big(-{275913\over4456448} \: 
y^2+{7\over512}\Big) \: y^2-{7\over 416}\Big] \: y^2+{15\over704}\Big\}
\: y^2-{1\over36}\bigg) \: y^2\nonumber
\\ & & +{1\over28}\bigg] \: y^7
\end{eqnarray}

\subsection{Small degeneracy ($\eta<0$) \cite{Cox, Pichon}}

Define $\tilde{K}_\nu(x) = \e^x K_\nu (x)$ where $K_\nu$ is the modified
Bessel function, then:

\begin{equation}\label{fo:ND1}
F_{1\over2} (\eta, \beta)={1\over \sqrt{2\beta}} \sum_{n=1}^\infty {(-1)^{n-1}
\over n} \e^{n\eta} \tilde{K}_1 \left({n \over \beta}\right)
\end{equation}

\begin{equation}\label{fo:ND2}
F_{3\over2} (\eta, \beta)={1\over \sqrt{2\beta^3}} \sum_{n=1}^\infty 
{(-1)^{n-1} \over n} \e^{n\eta} \left[\tilde{K}_2 \left({n \over \beta}
\right)-\tilde{K}_1 \left({n \over \beta}\right)\right]
\end{equation}

\begin{equation}\label{fo:ND3}
F_{5\over2} (\eta, \beta)={1\over \sqrt{2\beta^5}} \sum_{n=1}^\infty 
{(-1)^{n-1} \over n} \e^{n\eta} \left\{2\left[\tilde{K}_1 \left({n \over 
\beta}\right)-\tilde{K}_2 \left({n \over \beta}\right)\right]+{3\beta 
\over n} \tilde{K}_2\left({n \over \beta}\right)\right\}
\end{equation}

While for $\eta>-30$ and $\log \beta<-1$, the following series expansions
by Miralles \& Van Riper \cite{Mira} are better:

\begin{equation}\label{fo:SD1}
F_{1\over2} (\eta, \beta)={\sqrt{\pi}\over 2} \e^\eta \left(1+{3\over8} \beta
-{15\over128}\beta^2+{105\over1024} \beta^3-{105\over1024}\beta^4\right)
\end{equation}

\begin{equation}\label{fo:SD2}
F_{3\over2} (\eta, \beta)={3\sqrt{\pi}\over 4} \e^\eta \left(1+{5\over8} \beta
-{35\over128}\beta^2-{2345\over16384} \beta^3\right)
\end{equation}

\begin{equation}\label{fo:SD3}
F_{5\over2} (\eta, \beta)={\sqrt{\pi}\over 2} \e^\eta \left(1+{7\over8} \beta
-{539\over4090}\beta^2\right)
\end{equation}

\subsection{Global fit from EFF \cite{Egg, Pichon}}

If one changes the parameters of the generalized FD functions from ($\eta, 
\beta$) to ($f, g$) where:

\begin{equation}\label{fo:EFF1}
\eta = 2 u +\ln \left( {u-1 \over u+1} \right)
\end{equation}

\begin{equation}\label{fo:EFF2}
\beta = g / u 
\end{equation}

and

\begin{equation}\label{fo:EFF3}
u = \sqrt{1+f} 
\end{equation}

then the generalized FD functions can be approximated as

\begin{equation}\label{fo:EFF4}
F_{1 \over 2} (\eta, \beta) = (\sqrt{2} \beta^{3/2})^{-1} \: (\hat{\rho} 
+\hat{u}-3\hat{p})
\end{equation}

\begin{equation}\label{fo:EFF5}
F_{3 \over 2} (\eta, \beta) = (\sqrt{2} \beta^{5/2})^{-1} \: (3\hat{p} 
-\hat{u})
\end{equation}

\begin{equation}\label{fo:EFF6}
F_{5 \over 2} (\eta, \beta) = (\sqrt{2} \beta^{7/2})^{-1} \: (2\hat{u} 
-3\hat{p})
\end{equation}

where

\begin{equation}\label{fo:EFF7}
\hat{\rho}={f \over 1+f}\:g^{3/2}(1+g)^{3/2}\: {\sum^M_{m=0} \sum^N_{n=0} 
\hat{\rho}_{mn}f^m g^n \over (1+f)^M \: (1+g)^N}
\end{equation}

\begin{equation}\label{fo:EFF8}
\hat{p}={f \over 1+f}\:g^{5/2}(1+g)^{3/2}\: {\sum^M_{m=0} \sum^N_{n=0} 
\hat{P}_{mn}f^m g^n \over (1+f)^M \: (1+g)^N}
\end{equation}

\begin{equation}\label{fo:EFF9}
\hat{u}={f \over 1+f}\:g^{5/2}(1+g)^{3/2}\: {\sum^M_{m=0} \sum^N_{n=0} 
\hat{U}_{mn}f^m g^n \over (1+f)^M \: (1+g)^N}
\end{equation}

The values of the coefficients $\hat{\rho}_{mn}$, $\hat{P}_{mn}$ and 
$\hat{U}_{mn}$ can be found in \cite{Egg} with $M=N=1,2,3$ or $ 4$.

\section{APPENDIX: Proof of Eq. (\ref{fo:Cancel})}
  \label{Ap:Cancel}

We use the series expansion approximation to prove the following expression,
for which $k > -1$.

\begin{eqnarray}\label{fo:Cancel1}
   \{ [\partial F_k (\eta, \beta) / \partial \eta] + \beta
   [\partial F_{k+1}  (\eta, \beta) / \partial \eta] \}^2 \over
   [F_k (\eta, \beta) + \beta F_{k+1}  (\eta, \beta)] \: \:
   \{ [\partial^2 F_k (\eta, \beta) / \partial \eta^2] + \beta
   [\partial^2 F_{k+1}  (\eta, \beta) / \partial \eta^2] \}
 & & \nonumber \\
  \longrightarrow \left\{
   \begin{array}{ll}
   1 & \eta \rightarrow - \infty\\
   {k+1 \over k} & \eta \rightarrow \infty, \beta = 0, k \neq 0\\
   {k+5/2 \over k+3/2} & \eta \rightarrow \infty, \beta \neq  0
   \end{array}
 \right. & &
\end{eqnarray}

\subsection{Small degeneracy, non-relativistic}

When $\eta \rightarrow - \infty$, $\beta =0$, the approximation of the
generalized FD functions is given by Cox \& Giuli \cite{Cox} Eq. (24.111) as:

\begin{equation}
F_k (-\infty, 0) = F_k (-\infty) = \e^\eta \, \Gamma (k+1)
\end{equation}

So we get:

\begin{equation}
F_k (-\infty) = {\partial F_k (-\infty) \over \partial \eta} =
{\partial^2 F_k (-\infty) \over \partial \eta^2}
\end{equation}

Hence Eq. (\ref{fo:Cancel1}) can be written as:

\begin{equation}
{[ \partial F_k (-\infty) / \partial \eta]^2 \over F_k (-\infty)
[\partial^2 F_k (-\infty) / \partial \eta^2]} \longrightarrow \:\: 1
\end{equation}

\subsection{Small degeneracy, arbitrary relativistic}

When $\eta \rightarrow - \infty$, $\beta \neq 0$, we can find the leading term
of the generalized FD functions
for the following two extreme cases from \cite{Cox} Eq. (24.253) - (24.255) 
and Eq. (24.267) - (24.269), respectively, as:

\begin{equation}
\left\{ \begin{array}{llr} 
   F_k (-\infty, \beta) = \e^\eta g_k (\beta) & \beta \ll 1 \\
          F_k (-\infty, \beta) = \e^\eta h_k (\beta) & \beta \gg 1
\end{array} \right.
\end{equation}

For which we can see that:

\begin{equation}
F_k (-\infty, \beta) = {\partial F_k (-\infty, \beta) \over \partial \eta} =
{\partial^2 F_k (-\infty, \beta) \over \partial \eta^2}
\end{equation}

Hence Eq. (\ref{fo:Cancel1}) is satisfied.

\subsection{Great degeneracy, non-relativistic}

When $\eta \rightarrow \infty$, $\beta =0$ we get from \cite{Cox} Eq. (24.116) 
that

\begin{equation}
F_k (\infty, 0) = F_k (\infty) = {1 \over k+1} \; \eta^{k-1}
\end{equation}

So,

\begin{equation}
{[ \partial F_k (\infty) / \partial \eta]^2 \over F_k (\infty)
[\partial^2 F_k (\infty) / \partial \eta^2]} \longrightarrow \:\: 
{k+1 \over k}
\end{equation}

\subsection{Great degeneracy, arbitrary relativistic}

When $\eta \rightarrow \infty$, $\beta \neq 0$, the leading term of the 
generalized FD functions from \cite{Cox} Eq. (24.179) - (24.181) is:

\begin{equation}
F_k(\infty, \beta) = {1\over \sqrt{2}}\,\eta^{k+1}\, {f_k(y) \over 
(\sqrt{1+y^2}-1)^{k+1}}
\end{equation}

where $1+y^2=(1+\eta \beta)^2$. When $\eta \rightarrow \infty$, $\beta 
\neq 0$, $y \rightarrow \eta \beta \gg 1$. So, $\sqrt{1+y^2}-1 \rightarrow y$.
{}From \cite{Cox} Eq. (24.202), we can also find

\begin{equation}
f_k(y) \longrightarrow {1 \over k+3/2} \, y^{k+3/2}
\end{equation}

So,
\begin{equation}
F_k(\infty, \beta) \rightarrow {1\over \sqrt{2}}\,\eta^{k+1}\, {1 \over k+3/2}
{y^{k+3/2} \over y^{k+1}} = \sqrt{\beta \over 2} \, {1 \over k+3/2} \, 
\eta^{k+3/2}
\end{equation}

Hence we can see that

\begin{equation}
{[ \partial F_k (\infty, \beta) / \partial \eta]^2 \over F_k (\infty, \beta)
[\partial^2 F_k (\infty, \beta) / \partial \eta^2]} \longrightarrow \:\: 
{k+3/2 \over k+1/2}
\end{equation}

If we look at Eq. (\ref{fo:Cancel1}) we can see that when $\eta \rightarrow
\infty$ the $F_{k+1}(\infty, \beta)$ terms dominate over the $F_k(\infty,
\beta)$ terms. As a result, the ratio in Eq. (\ref{fo:Cancel1}) will be 
$k+5/2 \over k+3/2$.

\newpage

Table 1: Index of the function or derivative 
to be evaluated by the program.

\begin{center}
\begin{tabular}{|l c|l c|}
\hline \hline
{\rm IB} & Func. & {\rm IB} & Func. \\
\hline 
0 & $ F_k (\eta, \beta) $ & 
1 & ${\partial F_k (\eta, \beta) \over \partial \eta} $\\
2 & ${\partial F_k (\eta, \beta) \over \partial \beta}$ & 3 &
${\partial^2 F_k (\eta, \beta) \over\partial \eta^2}$\\
4 & ${\partial^2 F_k (\eta, \beta) \over\partial \eta \partial \beta}$ &
5 & ${\partial^2 F_k (\eta, \beta) \over\partial \beta^2}$\\
6 & ${\partial^3 F_k (\eta, \beta) \over \partial \eta^3}$ &
7 & ${\partial^3 F_k (\eta, \beta) \over \partial \eta^2 \partial \beta}$\\
8 & ${\partial^3 F_k (\eta, \beta) \over \partial \eta\partial \beta^2}$&
9 & ${\partial^3 F_k (\eta, \beta) \over \partial \beta^3}$\\
\hline 
\end{tabular}
\end{center}

\newpage

\begin{figure}
\includegraphics{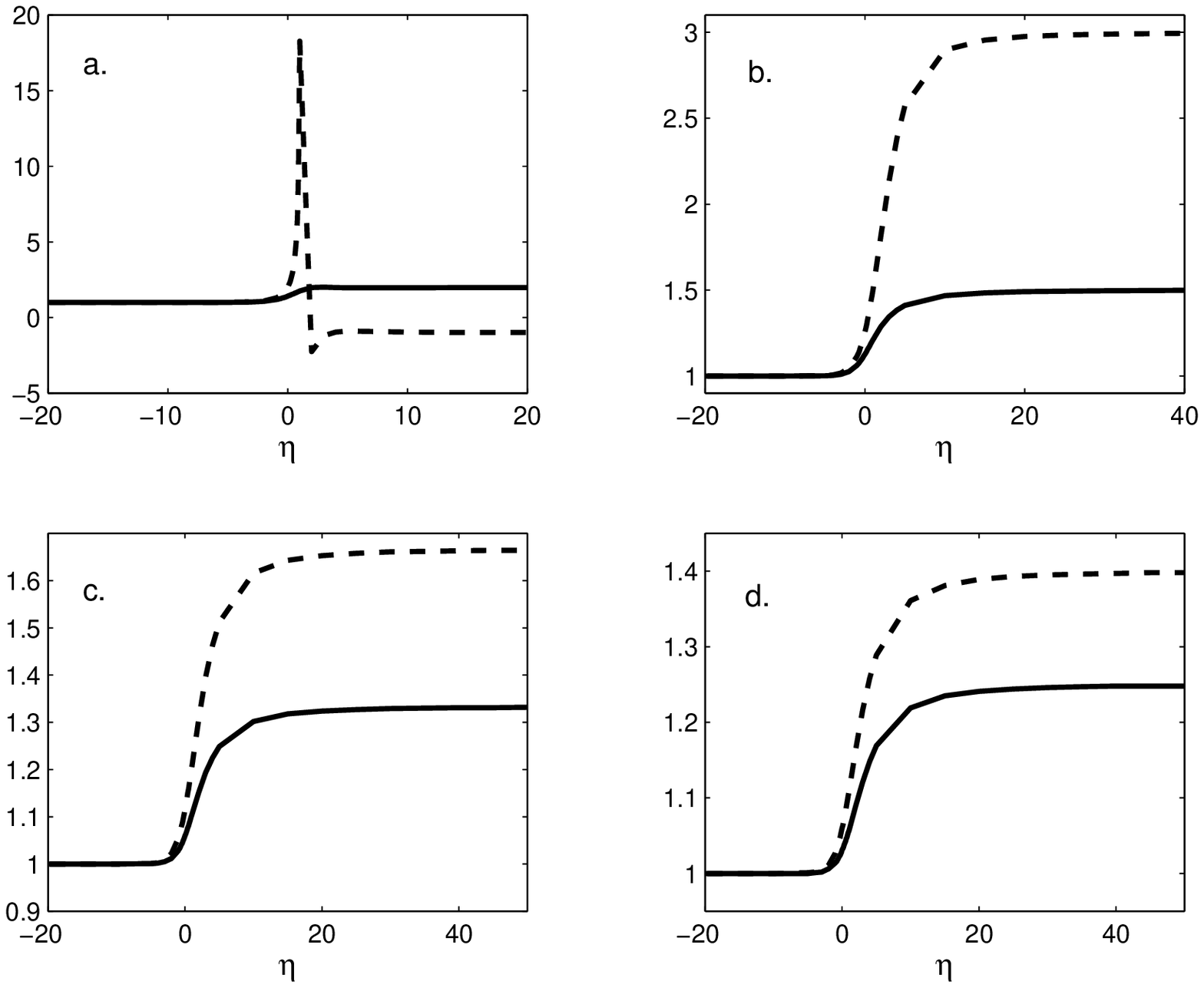}
\noindent{
Fig.~1: Different limits approached in Eq. (\ref{fo:Cancel}).  
   a. $k=-{1\over2}$, b. 
   $k={1\over2}$, c. $k={3\over2}$, d. $k={5\over2}$. Solid line for
   $\beta=1$, and dashed line for $\beta=0$.
}
\end{figure}

\begin{figure}
\noindent{
Fig.~2: Behavior of the tuning functions for $\beta=1$ and
   $\eta=10$ case. Indexes refer to the corresponding derivatives in
   Table (\ref{tb:Index}). 
}
\includegraphics{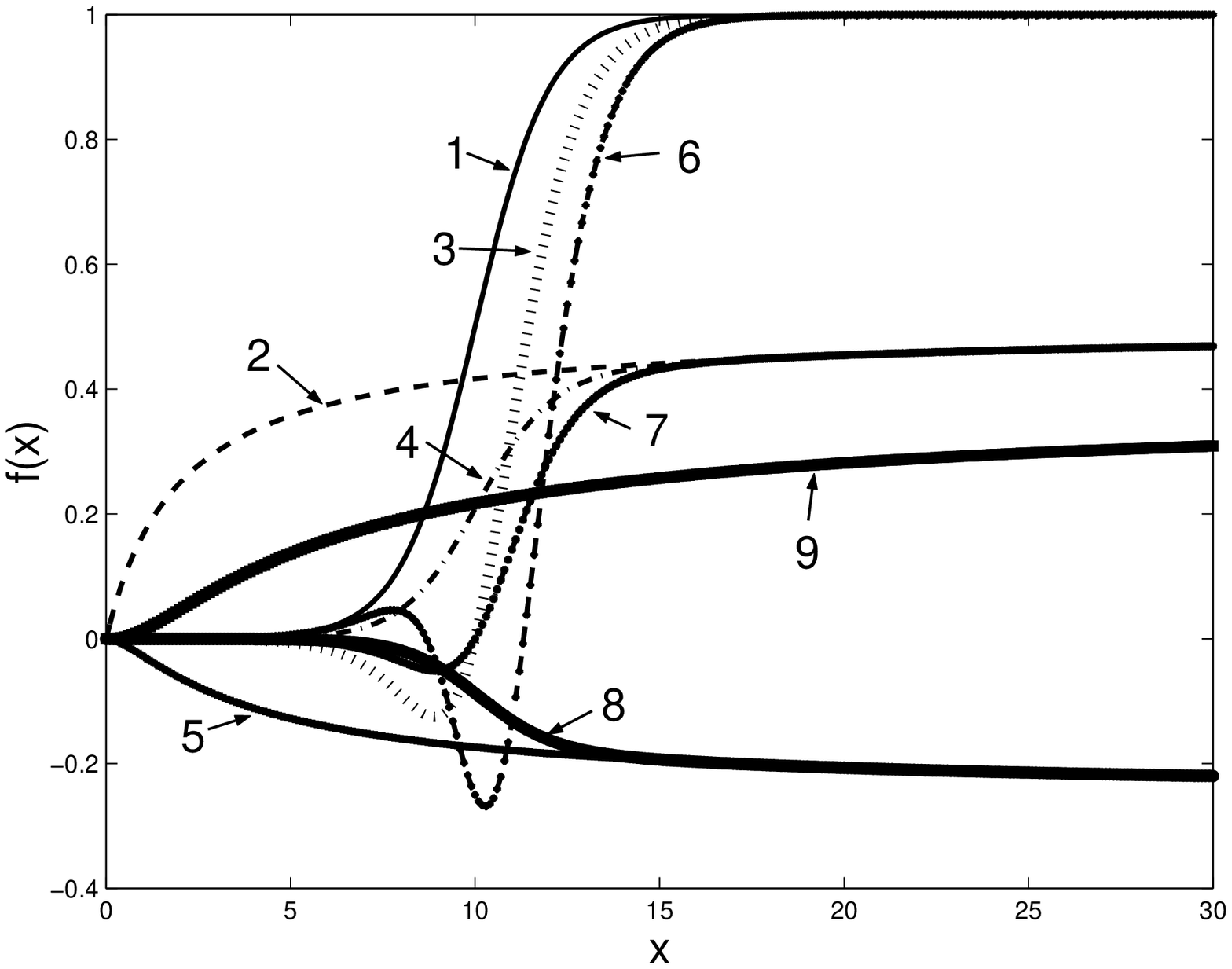}
\end{figure}

\begin{figure}
\noindent{
Fig.~3: Effective range of non-relativistic approximation for
  $F_{1\over2} (\eta, \beta)$ from Eq. (\ref{fo:NR}). $y$ axis is the
  logarithm of relative difference $\log (|F_{approx}-F|/F)$; the same
  notation for Figs. (\ref{fig:NR1}) to (\ref{fig:Egg}).
}
\includegraphics{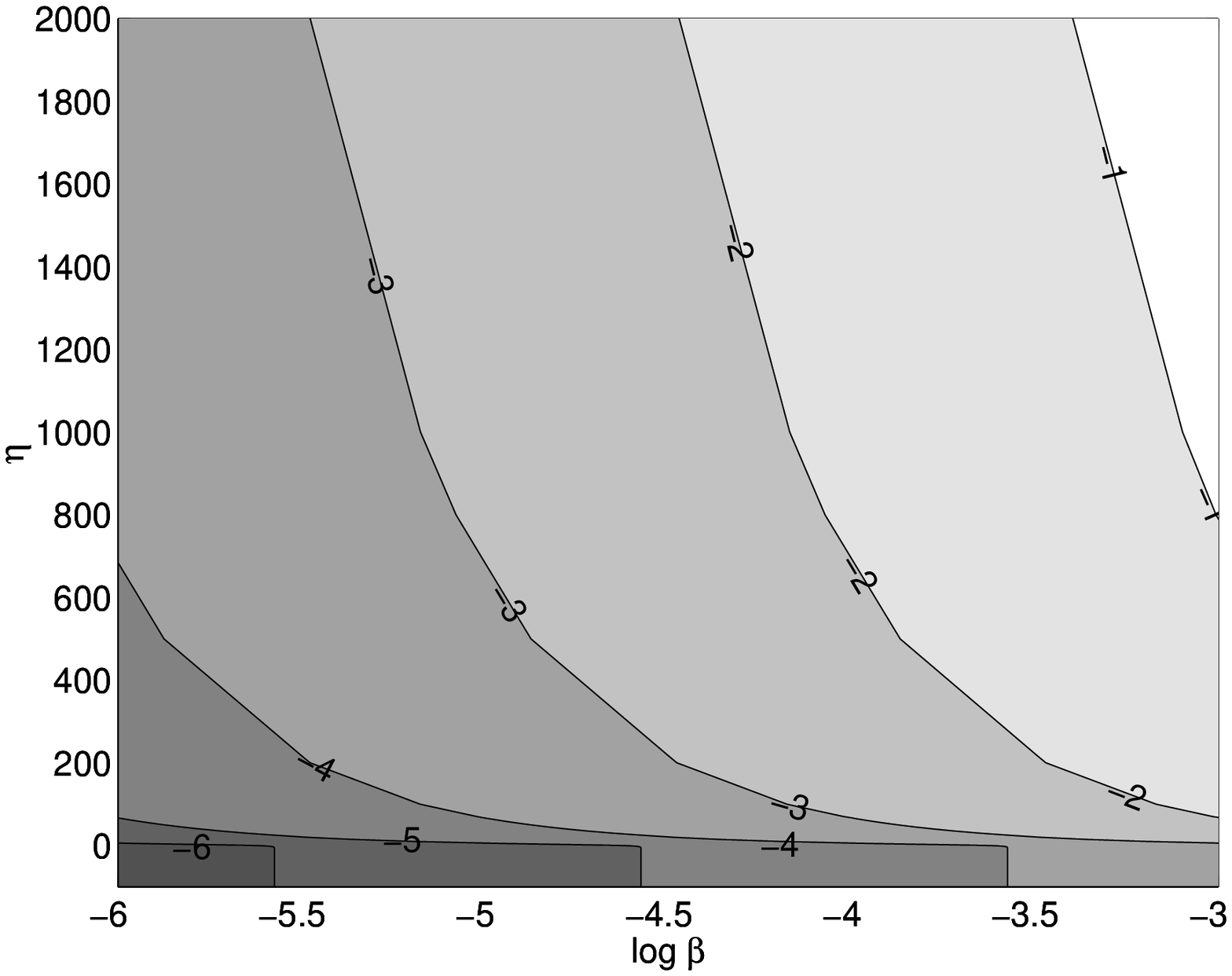}
\end{figure}

\begin{figure}
\noindent{
Fig.~4: Effective range of non-relativistic approximation for
  $F_{1\over2} (\eta, \beta)$ from Eq. (\ref{fo:NR1}). Notations are 
  the same as in Fig. (\ref{fig:NR}).
}
\includegraphics{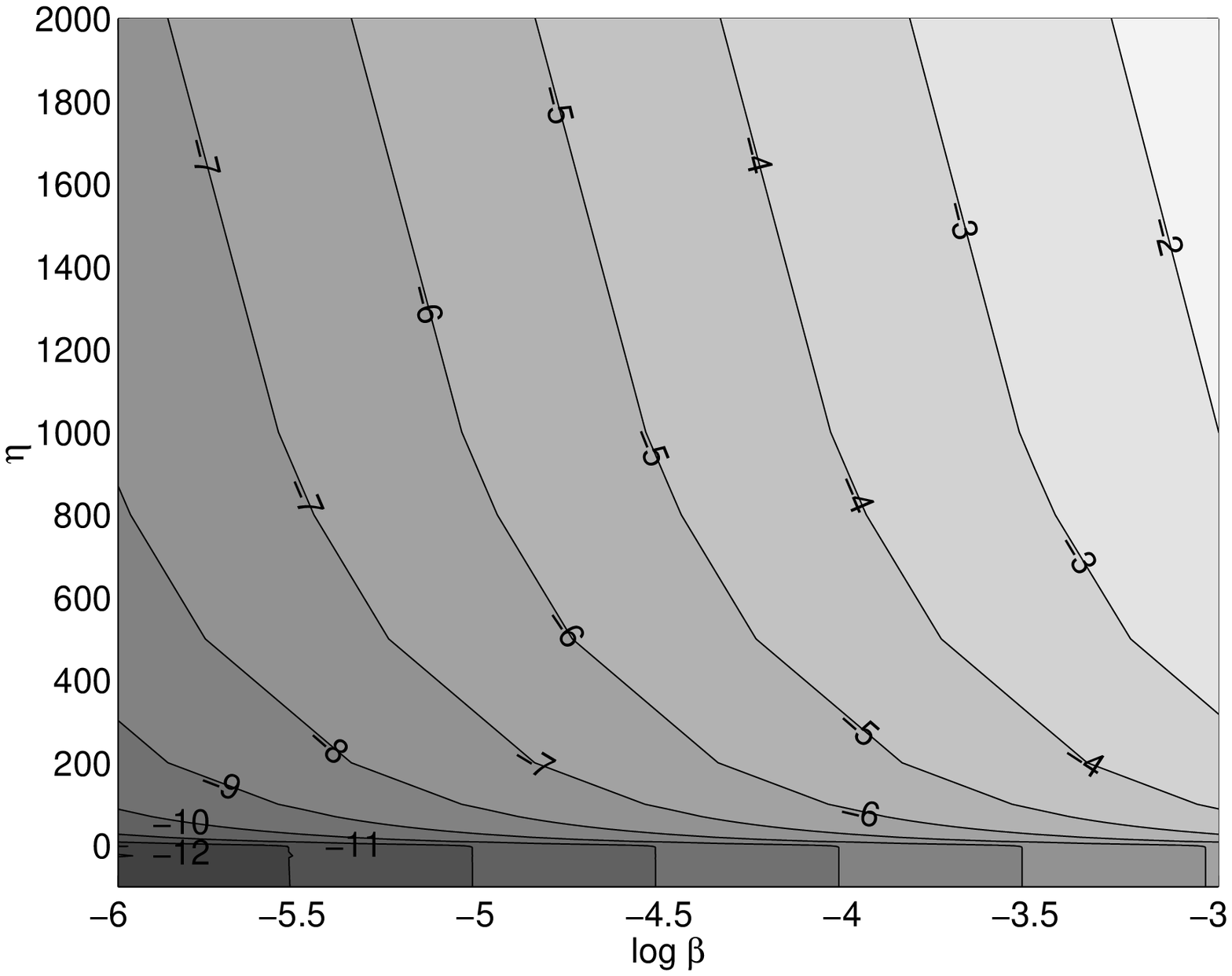}
\end{figure}

\begin{figure}
\noindent{
Fig.~5: Effective range of ultrarelativistic approximation for
  $F_{1\over2} (\eta, \beta)$ from Eq. (\ref{fo:ER}). Notations are 
  the same as in Fig. (\ref{fig:NR}).
}
\includegraphics{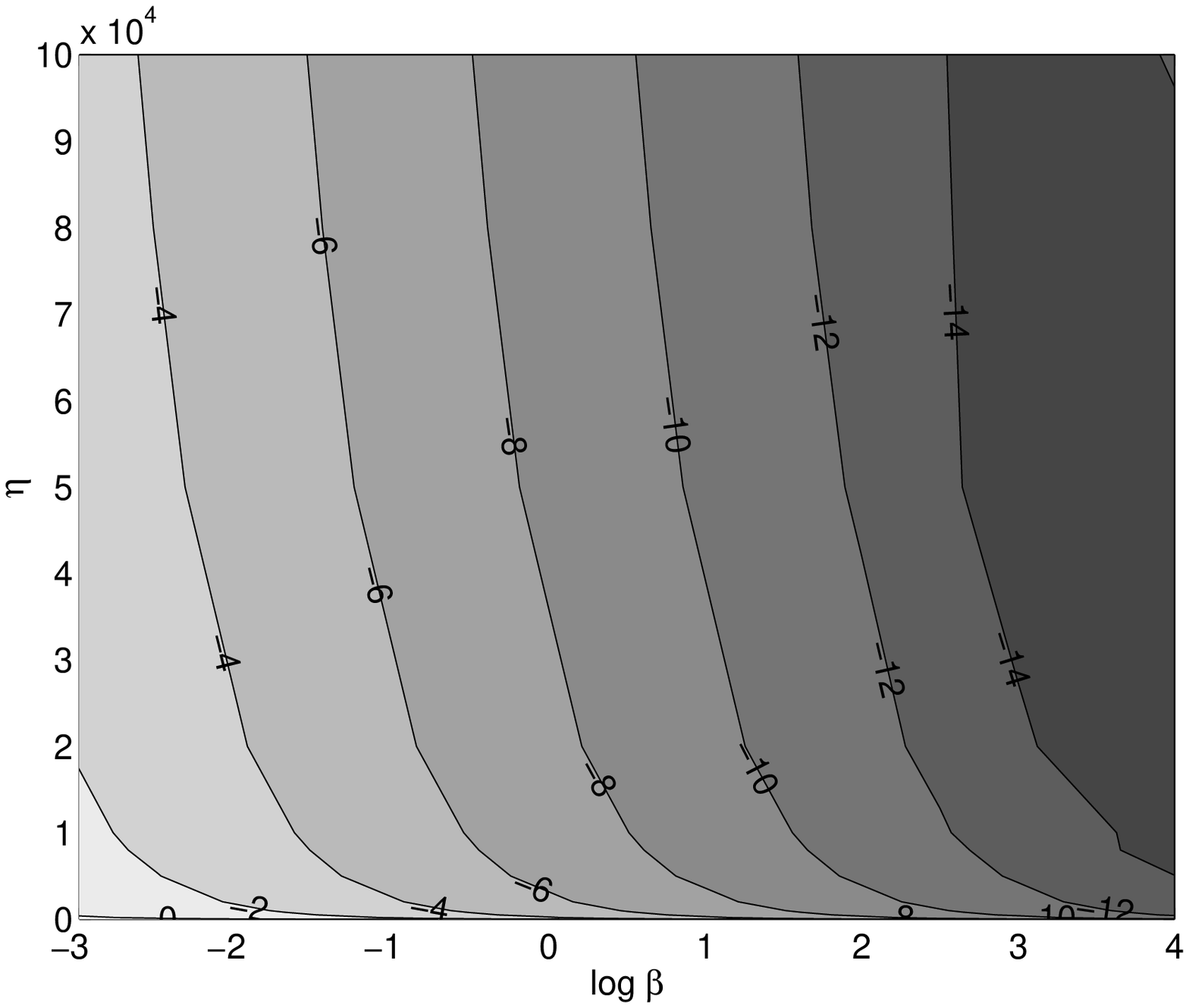}
\end{figure}

\begin{figure}
\noindent{
Fig.~6: Effective range of extremly-degenerate approximation for
  $F_{1\over2} (\eta, \beta)$ from Eq. (\ref{fo:ED1}). Notations are 
  the same as in Fig. (\ref{fig:NR}).
}
\includegraphics{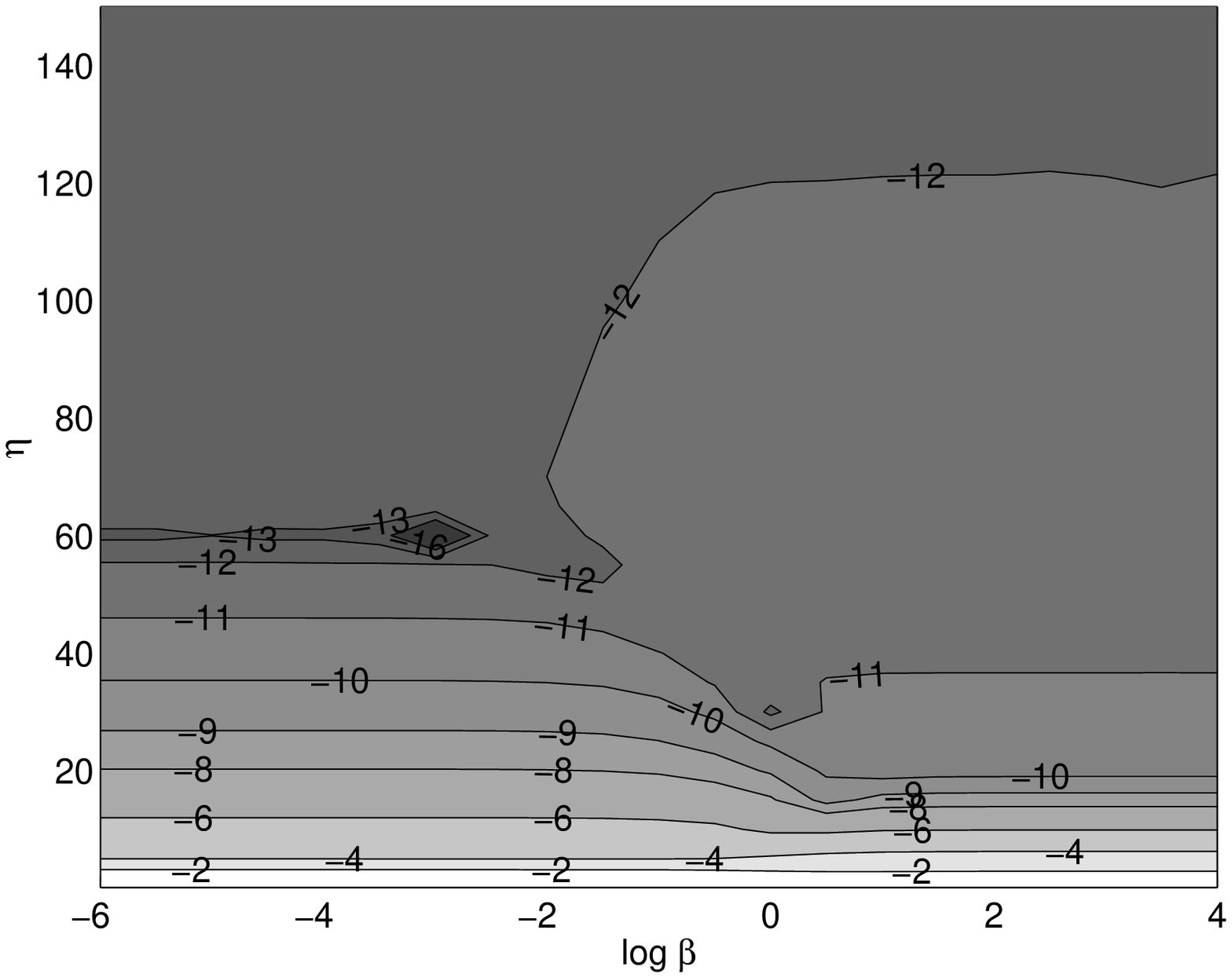}
\end{figure}

\begin{figure}
\noindent{
Fig.~7: Effective range of extremly-degenerate approximation for
  $F_{3\over2} (\eta, \beta)$ from Eq. (\ref{fo:ED2}). Notations are 
  the same as in Fig. (\ref{fig:NR}).
}
\includegraphics{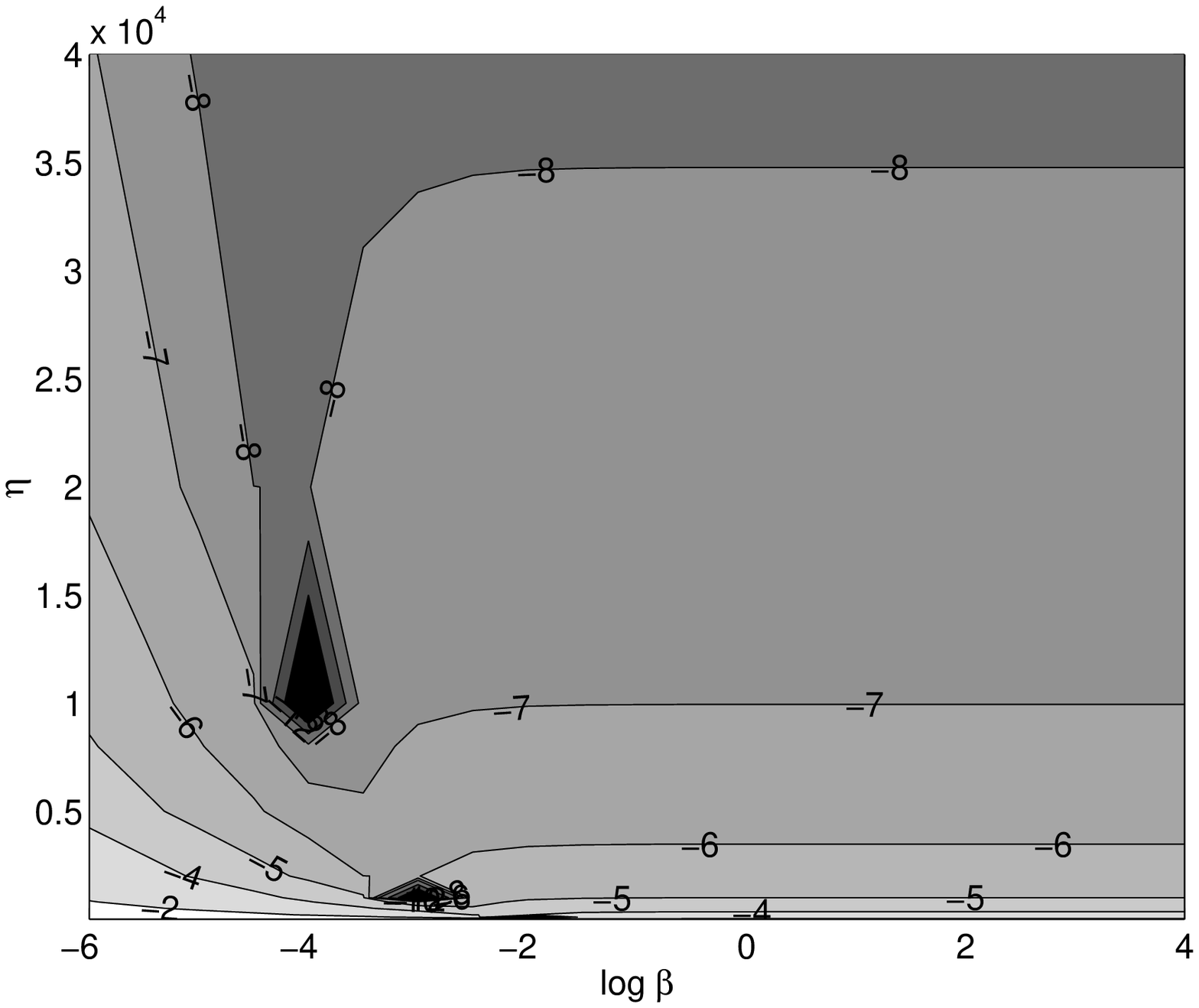}
\end{figure}

\begin{figure}
Fig.~8: Effective range of non-degenerate approximation for
\noindent{
  $F_{1\over2} (\eta, \beta)$ from Eq. (\ref{fo:ND1}). Notations are 
  the same as in Fig. (\ref{fig:NR}).
}
\includegraphics{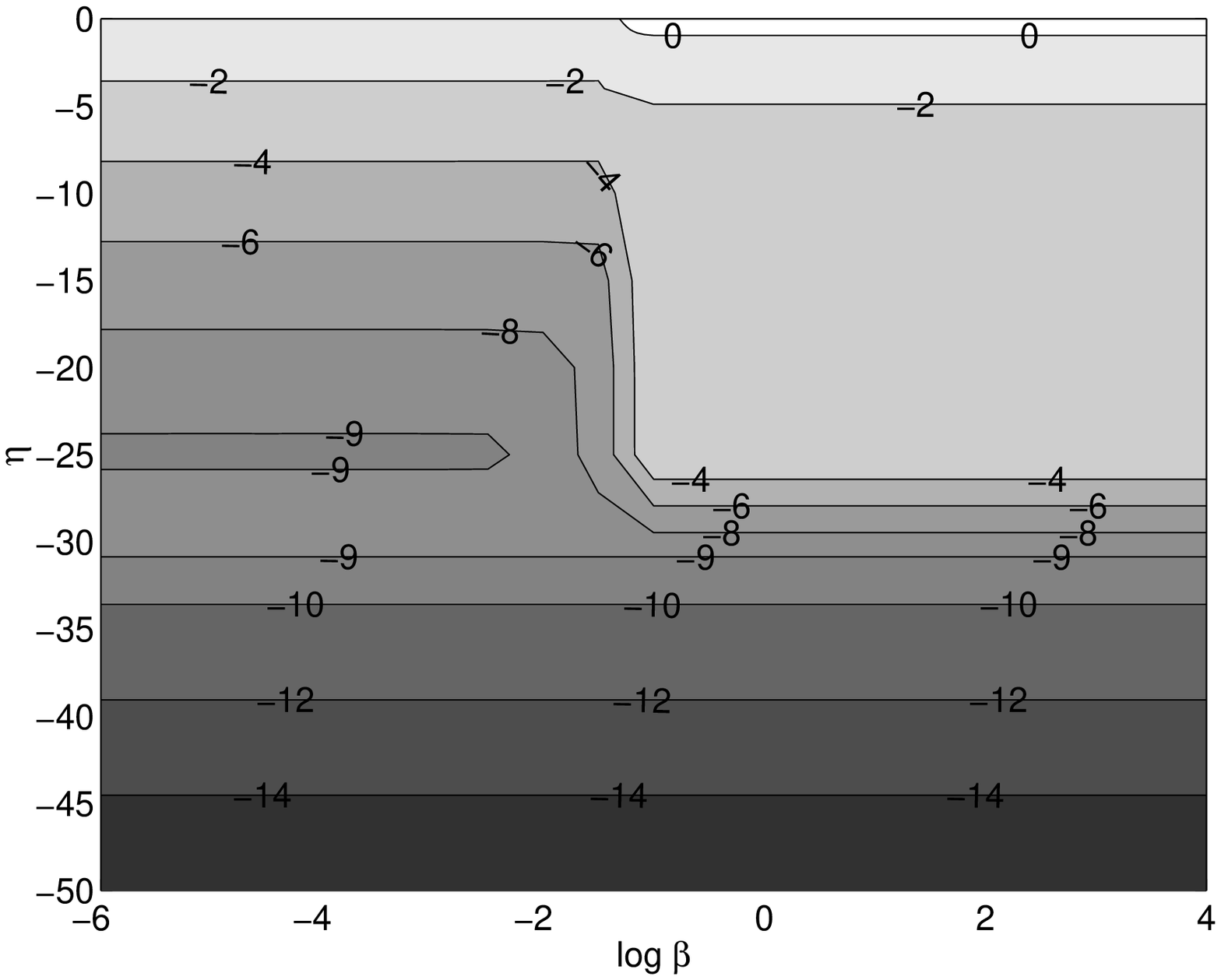}
\end{figure}

\begin{figure}
\noindent{
Fig.~9: Effective range fitting formula of $F_{1\over2} 
  (\eta, \beta)$ by \cite{Egg} (see \cite{Pichon}). Notations are the 
  same as in Fig. (\ref{fig:NR}).
}
\includegraphics{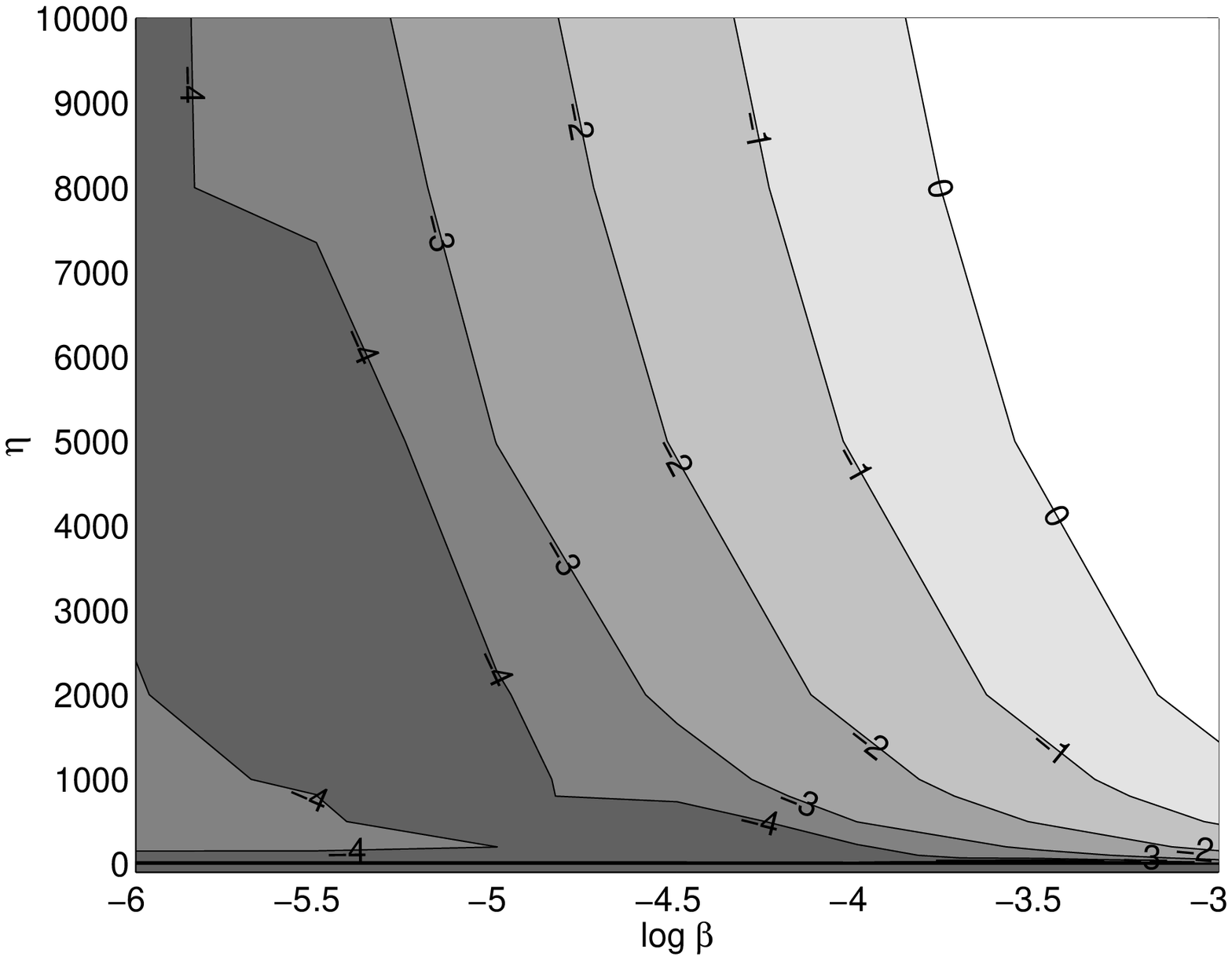}
\end{figure}

\end{document}